# Whole Cross-Sectional Human Ultrasound Tomography

David C. Garrett†, Jinhua Xu†, Yousuf Aborahama, Geng Ku,
Konstantin Maslov, and Lihong V. Wang*

*Caltech Optical Imaging Laboratory, Andrew and Peggy Cherng Department of Medical Engineering, Department of Electrical Engineering, California Institute of Technology, Pasadena, CA 91125 USA*

† These authors contributed equally.
* Corresponding author: lvw@caltech.edu

**Abstract**

Ultrasonography is a vital component of modern clinical care, with handheld probes routinely used for diagnostic imaging and procedural guidance. However, handheld ultrasound imaging is limited by factors such as the partial-cross-sectional field of view, operator dependency, contact-induced distortion, and lack of transmission contrast. Here, we demonstrate a new system enabling whole cross-sectional ultrasound tomography of humans in reflection and transmission modes. We generate 2D images of the entire *in vivo* human cross-section with uniform in-plane resolution using a custom 512-element circular ultrasound receiver array and a rotating ultrasonic transmitter. We demonstrate this technique in regions such as the abdomen and legs in healthy volunteers. To address unmet clinical needs, we explore two key applications. First, we readily observe abdominal adipose distributions in our images, enabling adipose thickness assessment over the body without ionizing radiation or mechanical deformation. Second, we demonstrate an approach for video-rate (30 frame-per-second) biopsy needle localization with respect to internal tissue features. These capabilities make whole cross-sectional ultrasound tomography a potential practical tool for clinical needs not currently met by other modalities.

**Introduction**

Since its inception in the mid-20th century, ultrasound imaging has revolutionized healthcare by enabling non-invasive and affordable visualization of soft tissue structure and function. Early systems employed single transducers scanned linearly or circularly with subjects immersed in a water bath [1], [2], later followed by membranes or articulating arms to image regions in the abdomen [3], [4]. Initial results were promising for disease diagnosis [5], but imaging required mechanical scanning over ~1 hour [6]. Later developments in transducers and electronics led to linear probes [7], where multiple channels could be used in parallel. These developments led to the modern handheld probe, which has become the predominant form of ultrasonography across numerous clinical applications. However, probes require trained operation [8], have limited ability to visualize features behind bone or air pockets, and provide only reflection-mode images over a narrow field of view (FOV). The FOV can be expanded by scanning the probe around the periphery of regions like the human thigh and co-registering adjacent frames [9], [10]. However, existing approaches require manual probe movement to maintain contact with the skin, which can lead to image variation between operators and systems [11], [12].

More recently, alternate approaches using smaller immersion tanks with planar [13], linear [14], ring [15], or hemispherical [16] transducer arrays have been investigated for ultrasound tomography (UST) imaging of the breast [17] or limbs. These systems record both reflected and transmitted signals, allowing for the generation of reflectivity, speed of sound, and attenuation coefficient profiles. The extension to human-scale imaging has been historically constrained by acoustic propagation barriers, particularly at bone and air interfaces, where extreme impedance mismatches hinder wave transmission. Nevertheless, a recent study achieved whole-cross-sectional imaging of ~12 cm diameter piglets despite the presence of bone and air [18]. Although this method produces high-quality images, it requires substantial data acquisition time (15 – 25 minutes) and reconstruction time (20 – 25 minutes). Such extended protocols are challenging to adapt to *in vivo* human imaging due to subject motion. These times also rival those of MRI, which may limit the practicality and scalability of the technique in clinical settings. Another recent system enables volumetric reflection-mode imaging of vasculature and bones in human extremities like the arm [19]. However, both of these systems' parameters (e.g., array dimensions, acoustic frequency and power, and detection sensitivity) are not yet suitable for human-scale imaging in regions like the abdomen.

In this work, we developed a system that enables UST of the entire human cross-section, resulting in 2D images of reflectivity, speed of sound, and attenuation coefficient profiles. We constructed a custom 512-element circular receiver array combined with a single-element transmitter that rotates around the subject. This geometry enables both reflection- and transmission-mode imaging with improved robustness to acoustic occlusions, while maintaining practicality for *in vivo* human abdominal imaging compared to planar or hemispherical alternatives. To image deep in the body, we enhance the signal sensitivity by using low-noise parallel preamplifiers directly coupled with the receiver array and by exciting the transmitter with a chirp waveform. Compared to handheld probes, we reduce issues of acoustic shadowing from regions containing bone or air pockets by using full 360° viewing angles. In comparison to MRI and other standard imaging modalities, whole cross-sectional UST is a potential low-cost, safe, fast, and convenient tool for screening and monitoring abdominal conditions.

We demonstrate this technique by imaging the abdomen and legs of healthy volunteers, where we clearly observe several organs and key features in reflection-mode images, and we obtain profiles of speed of sound and attenuation coefficient. This approach visualizes abdominal adipose layers, offering a promising modality for mapping adipose thickness distributions without ionizing radiation or tissue compression. We also demonstrate an approach for localizing biopsy needle tips deep in tissue with respect to internal features. By coupling an acoustic transmitter to a commercial needle and detecting the resulting scattered signals from the needle tip, we obtain 30 frame-per-second images of the needle tip location. Together, these techniques showcase whole cross-sectional UST as a safe and practical modality for a variety of clinical applications.

**Results**

*Whole cross-sectional human imaging*

We developed a custom 60 cm diameter, 512-element circular acoustic receiver array with 1 MHz center frequency. A 1.5-inch diameter transducer (Olympus V395) with a custom cylindrical diverging polymethylpentene (TPX) lens is used as a transmitter, and it is mounted on a plastic gear that rotates around the subject using a stepper motor. All receiver channels are preamplified using custom circuit boards inside the array (Supplementary Fig. 1), and these signals are digitized in parallel using two data acquisition modules (DAQs, Photosound Legion) inside shielded enclosures. The array is mounted on two vertical motor stages to adjust its height in a water immersion tank. Water acts as acoustic coupling between the skin and transducers. An arbitrary function generator (Siglent SDG2042X) connected to a 300-Watt RF power amplifier (ENI A300) excites the transmitter. The system hardware is shown in Fig. 1a and b.

To enhance the signal-to-noise ratio (SNR) without exceeding the mechanical index (MI) safety standard, we use a 400 μs chirp signal spanning 0.3 – 2.0 MHz (Fig. 1c). We first record the transducer response using only water in the imaging domain, which is then cross-correlated with the target response to recover a pulse-like representation. We display example signals from the receiver array for water and the human abdomen in Fig. 1d, showing both backscattered and transmitted signals recorded in parallel. Fig. 1e shows individual traces for the channel opposite the transmitter (channel 256). This approach also enables channel calibration using the water scan, where we expect the same received amplitude for channels directly opposite the transmitter.

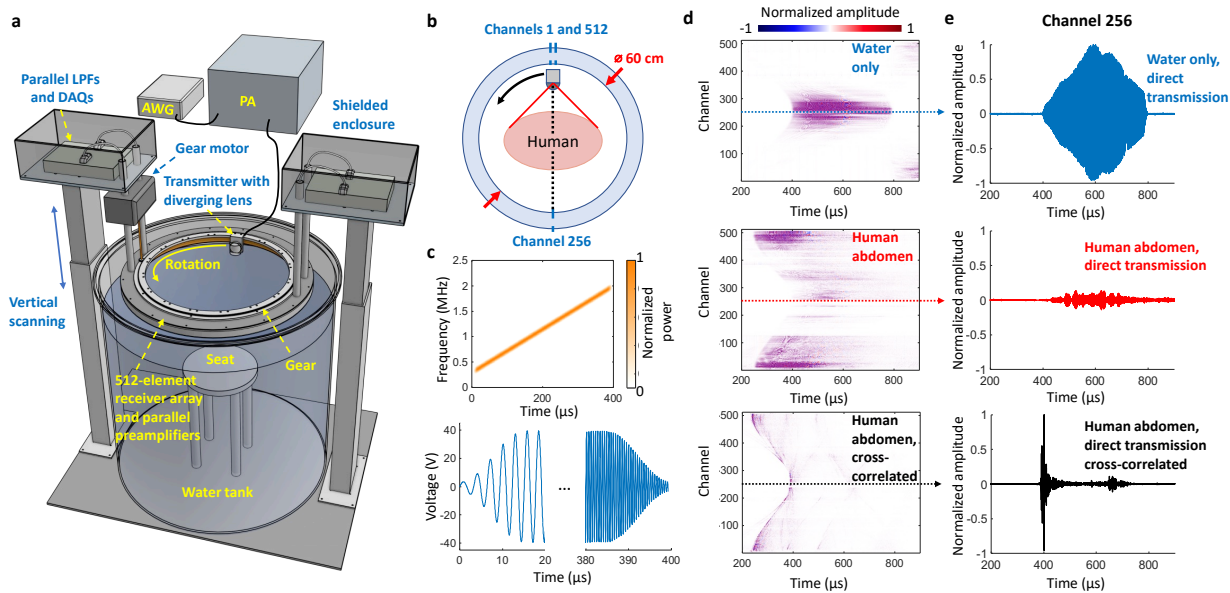

**Fig. 1. Whole cross-sectional UST system. a** System diagram. AWG: arbitrary waveform generator; PA: power amplifier; LPFs: low-pass filters; DAQs: data acquisition modules. **b** Schematic diagram of the transmitter and array, showing example receiver array channel numbers. The dotted line shows the path of direct transmission through the body. **c** Chirp excitation waveform. Top: spectrogram of the 400 μs chirp spanning 0.3 – 2.0 MHz. Bottom: temporal profile of the amplified chirp signal coupled to the transmitter, with Hamming window apodization at the beginning and end of the signal. **d** Example signals recorded with the receiver array. Top: water only. Middle: human abdomen. Bottom: human abdomen after cross-correlation with the water only signals. The dashed line indicates channel 256 (shown in panel **e**). **e** Example signals from an individual receiver channel directly opposite the transmitter.

We demonstrate cross-sectional UST with healthy volunteers. For abdominal imaging, subjects sit on a stool in the water immersion tank with their head supported against a cushion to reduce motion and with their arms raised slightly to elevate the rib cage. During a 10-second scan, subjects are asked to hold their breath while remaining still. Fig. 2a shows an example reflection-mode abdominal image of a 24-year-old female volunteer. The image is displayed in inverse grayscale (brighter regions are more anechoic) normalized to the peak pixel amplitude. The outer boundary is extracted using an automated segmentation tool [20]. These images visualize various structures, such as the liver, stomach, spleen, abdominal aorta, and vertebral body. We labeled regions using an anatomical atlas and through consultation with a clinical collaborator. Note that despite the presence of bone and potential air pockets, our geometry enables imaging of regions deep in the body. Due to our lower acoustic frequency than typical probe-based ultrasonography, our images correspond primarily to reflections from tissue boundaries rather than from scattering within tissues [21]. With the same volunteer standing in the immersion tank, we also imaged the legs as shown in Fig. 2b. In the upper legs, the femur, surrounding muscle groups, and adipose boundaries are observed. We visualize the tibia and fibula in the lower legs as well as adipose boundaries.

The signals transmitted through the body enable reconstruction of speed of sound and attenuation coefficient profiles, which are overlaid on the reflection-mode images in Fig. 2c and d. We observe a higher speed of sound in the liver compared to other organs, consistent with literature values [22], while greater attenuation coefficient is observed in the spine and stomach. Since we obtain distinct organ boundaries from the reflection-mode images, we can also segment organ contours to constrain the transmission-mode reconstruction for bulk organ regions as shown in Fig. 2e. We expect negligible transmission through bones like the vertebral body, so we do not solve for their speed of sound. We otherwise find good agreement with literature values for various organs such as the liver (Supplementary Fig. 6). We validated this approach with reference ethanol-water mixtures, simultaneously imaging five targets of ~3 cm diameter with speed of sound ranging from 1510 m/s to 1610 m/s. Our speed of sound estimations show less than 3% error (Supplementary Table 4), supporting this approach as a potential tool for quantitatively evaluating conditions like liver fibrosis or non-alcoholic fatty liver disease [23].

With another female volunteer, we performed scans at 1 cm vertical intervals from approximately the ribcage to the pelvis. The subject was in the immersion tank for approximately 10 minutes over the entire imaging session. Fig. 3 shows example 2D images. We observe additional features in some slices, such as the pancreas, hepatic portal vein, and kidneys. The entire liver cross-section is clearly visualized, highlighting the potential of cross-sectional UST for liver health assessment.

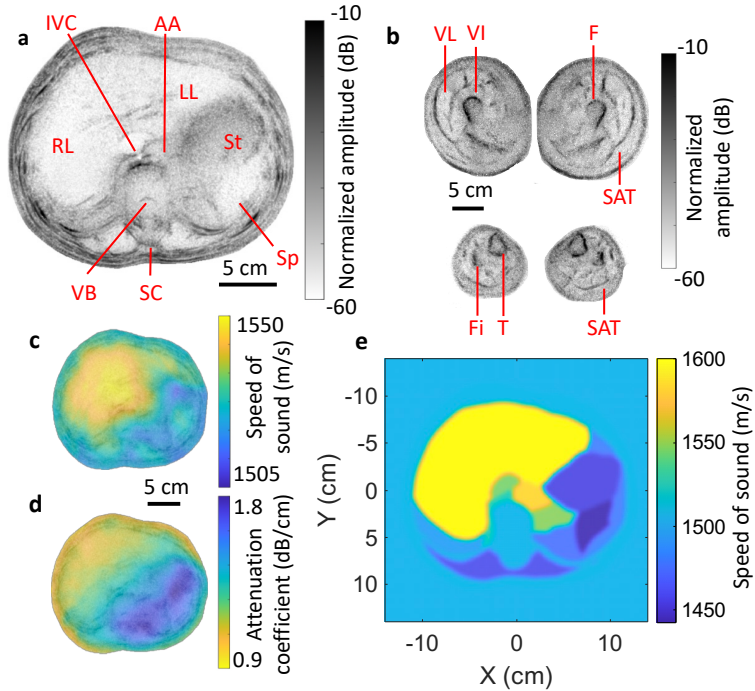

**Fig. 2. Whole cross-sectional UST of a healthy female. a** Reflectivity image of human abdomen. IVC: inferior vena cava. AA: abdominal aorta. RL: right lobe of liver. LL: left lobe of liver. VB: vertebral body. SC: spinal cord. St: stomach. Sp: spleen. **b** Reflectivity image of human upper leg (top) and lower leg (bottom). F: femur. Fi: fibula. T: tibula. VL: vastus lateralis. VI: vastus intermedius. SAT: subcutaneous adipose tissue. Panels **c** and **d** show the speed of sound and attenuation coefficient profiles, respectively, overlaid on the reflectivity image. **e** Speed of sound reconstruction constrained using organ regions determined from the reflectivity image.

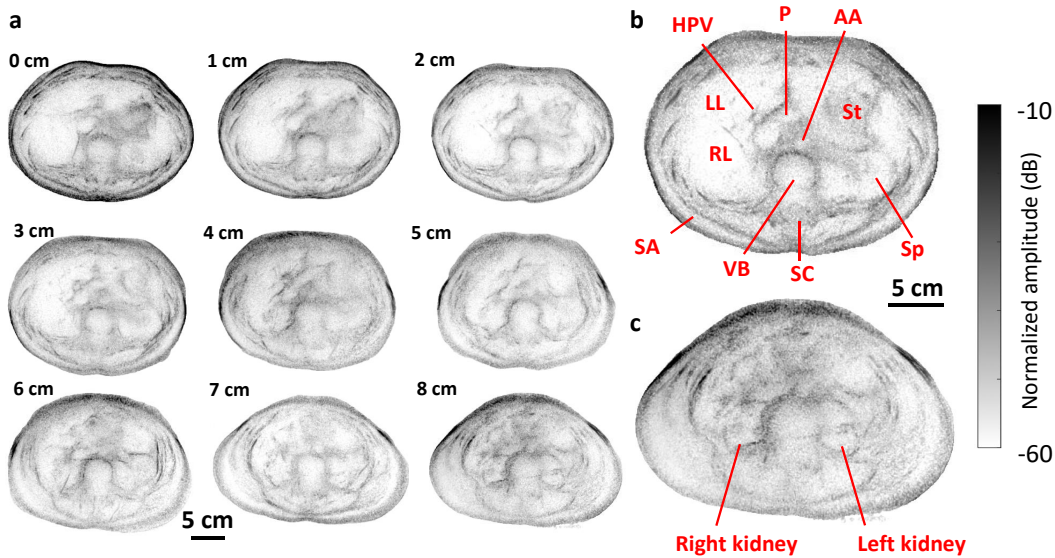

**Fig. 3. Example reflection-mode UST images of a healthy female's abdomen. a** Example images, with labels denoting the distance downward from the ribcage. **b** Expanded view of the 3 cm image. HPV: hepatic portal vein. IVC: inferior vena cava. AA: abdominal aorta. RL: right lobe of liver. LL: left lobe of liver. St: stomach. SA: subcutaneous adipose. VB: vertebral body. SC: spinal cord. Sp: spleen. **c** Expanded view of the 8 cm image.

*Adipose thickness assessment*

Subcutaneous adipose (SA) and preperitoneal adipose (PA) distributions serve as key indicators of metabolic health [24]. SA thickness can be measured with calipers and probe-based ultrasound. Calipers are known to be less accurate for individuals with larger SA thickness, are operator-dependent, and have only moderate agreement with estimates from MRI [25]. Probe-based ultrasound, though more accurate and capable of measuring PA thickness than calipers [26], requires extensive operator training and is affected by probe-induced tissue compression [27]. Both methods also require instrument repositioning at each measurement site, making frequent whole-body adipose distribution assessment impractical [28]. We compare existing methods with our approach in Supplementary Table 2.

Cross-sectional UST offers unique advantages for SA assessment by enabling clear visualization of the adipose layer around the abdominal periphery without mechanical deformation. To validate this approach, we first imaged a phantom consisting of a lard layer of varying known thicknesses held with plastic ribbon over a 4% agar core. We extracted the reflection-mode image amplitude along lines normal to the phantom surface, and we estimated the lard layer thickness by determining the distance between the two dominant peaks along this line (Supplementary Fig. 10). Our estimated thicknesses of $1.04 \pm 0.02$ cm and $2.02 \pm 0.05$ cm agree well with the true values of 1.00 cm and 2.00 cm, respectively.

We then imaged a healthy subject (25-year-old female) to compare our UST SA thickness estimates with caliper measurements. As shown in Fig. 4a and b, our adipose thickness estimation from the extracted line profile ($1.32 \pm 0.06$ cm) agrees closely with the caliper measurement of 1.3 cm used at the same location. We then imaged a 27-year-old male volunteer with greater SA thickness (Fig. 4c). The anterior side of the abdomen clearly shows the SA and PA regions relative to the skin surface and rectus abdominus muscles. Within the SA, we observe two distinct layers: the superficial and deep adipose layers, separated by the superficial (or Scarpa's) fascia [29], [30], [31], as shown in Fig. 4f. While internal features are less well visualized due to the greater tissue depth and potential increased air content in organs, the adipose layers remain clearly visible. To assess caliper accuracy in individuals with greater SA thickness, we obtained a UST image during caliper measurement. Fig. 4e demonstrates that calipers underestimate the total SA thickness [32], measuring 1.9 cm compared to approximately 3.2 cm in the UST image. Since whole cross-sectional UST is fast, safe, and more cost-effective than MRI, it shows promise for guiding and assessing weight loss regimes, clinical trials of weight loss drugs, tracking adipose malignancies (i.e., liposarcoma), or liposuction planning [33].

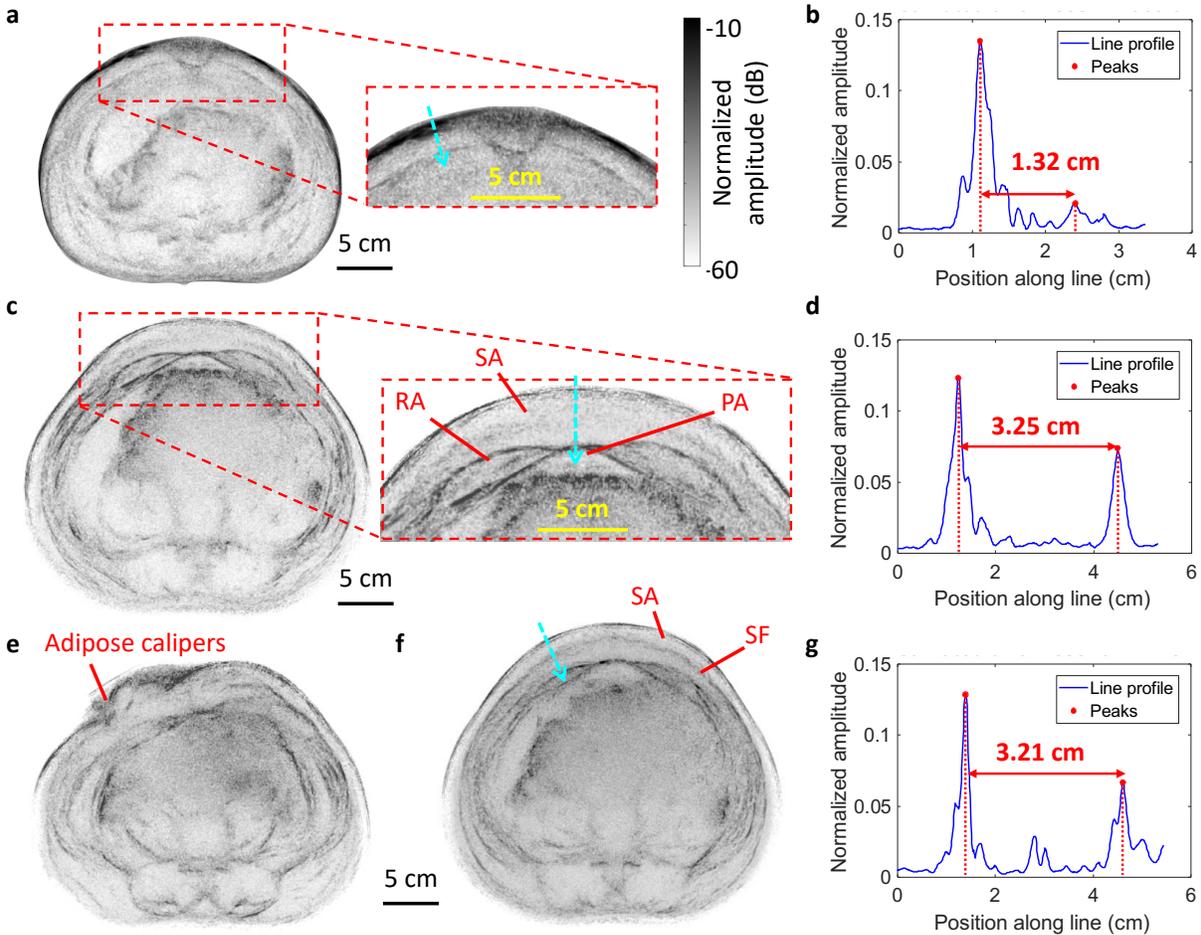

**Fig. 4. Abdominal adipose thickness assessment of healthy volunteers.** Panels **a** and **b** show UST results for a 25-year-old female, and panels **c** to **g** are of a 27-year-old male. **a** UST image of a female volunteer. **b** Image amplitude along the line drawn in **a** (in the direction of the arrow), estimating an adipose layer thickness of 1.32 cm near the navel. **c** UST image of the entire body of a male volunteer, with an inset showing adipose regions in the anterior region of the abdomen. RA: rectus abdominus. SA: subcutaneous adipose. PA: preperitoneal adipose. **d** Image amplitude along the line drawn in **c**. **e** UST image with adipose calipers positioned on the abdomen. **f** Abdominal image showing Scarpa's fascia (SF) in SA. **g** Image amplitude along the line drawn in **f**, where the adipose calipers were also positioned.

*Biopsy needle localization*

Next, we demonstrate UST-guided biopsy needle localization. In clinical practice, tissue samples from suspected cancerous regions are collected via needle biopsy, typically guided by ultrasound or X-ray CT imaging [34]. However, probe-based ultrasound localization is generally only used for superficial targets like in the breast, and it requires that the needle is approximately orthogonal to the imaging probe to provide sufficient backscatter to the probe. While manufacturers have developed treatments like scoring or bubble-filled polymer coatings to generate more isotropic scattering and improve the ultrasound visibility [35], these approaches can increase the insertional friction of the needle [36]. Moreover, probe-based localization requires the operator to simultaneously manipulate both the probe and needle, which requires substantial training [37] and becomes particularly challenging in complex tissue regions. CT needle guidance, while providing whole-body localization, involves a cumbersome workflow requiring multiple positioning cycles. Each adjustment requires the clinical team to exit and re-enter the scanning room, taking several minutes to complete. This interrupted workflow, combined with a concerning radiation exposure (approximately 10 mSv [38]), highlights the need for alternative guidance approaches.

Our approach uses a commercial 16-gauge core biopsy needle consisting of a solid stainless-steel core (1.5 mm diameter) that translates within a hollow sleeve. We found that ultrasound signals can be coupled into the unmodified needle using an ultrasonic transducer interfaced with the plastic handle. These signals propagate along the needle as an acoustic waveguide, and they are scattered nearly isotropically at the needle tip (Fig. 5a). To enhance SNR, we employ the same chirp signal used in UST imaging. The scattered signals from the tip are detected with our acoustic array and are cross-correlated with the chirp response. The propagation time along the needle is calibrated for and remains constant. We expect the dominantly excited mode to be longitudinal with approximately constant phase velocity over our frequency range (Supplementary Fig. 9).

The needle tip location is determined using one-way delay-and-sum beamforming, accounting for the propagation time along the length of the needle. The needle's image response (Fig. 5b) demonstrates a full width at half maximum (FWHM) of ~0.7 mm. This approach enables 30 frame-per-second needle localization over a human-scale FOV. Example video frames (Fig. 5c and d) show the needle response overlaid on a UST reflectivity image of an agarose phantom supported by a steel post. We normalize the needle response images by the same expected maximum value across all video frames. The needle's center response is automatically determined based on the 5$^{th}$ order moment of the needle image. The full video is given in Supplementary Video 1. As seen, the acoustic image quickly and accurately tracks the location of the needle tip with respect to the phantom, even when moving quickly or inserted into the tissue. To validate this approach in more realistic conditions, we used layered *ex vivo* porcine tissue (consisting of skin, adipose, and muscle) as an anatomical phantom. With the needle oriented approximately horizontally, this approach tracks its position during movement around and into the tissue (Fig. 5e and f). Supplementary Video 1 demonstrates continuous localization of the needle tip throughout the insertion path.

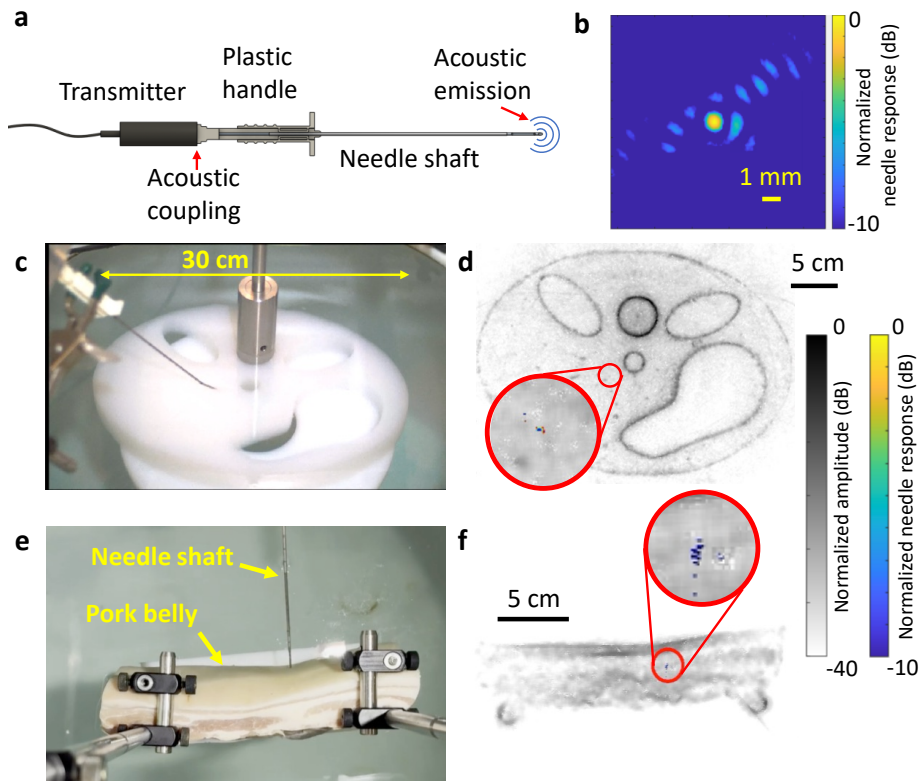

**Fig. 5. UST-guided biopsy needle localization. a** Diagram of needle configuration. **b** Representative image of the needle's acoustic response in water. **c** Video frame showing the needle inserted into an agarose phantom. **d** Reconstructed video frame overlaid on the reflectivity image. The red circle is automatically placed around the center acoustic response. **e** Video frame showing the needle inserted into *ex vivo* pork belly tissue. **f** Reconstructed video frame overlaid on the reflectivity image. The black and white and colored color bars on the right apply to the reflectivity and needle response images, respectively.

**Discussion**

We developed a system for whole cross-sectional human ultrasound imaging that advances beyond conventional probe-based ultrasonography in several key ways. Our approach captures complete body cross-sections while simultaneously providing three contrast mechanisms: reflectivity, speed of sound, and attenuation coefficient. Unlike clinical ultrasonography, which requires extensive operator training to visualize regions of interest, our approach could be largely automated since it requires minimal subject positioning. These advantages could be particularly attractive for applications requiring frequent imaging, and they could help reduce cost compared to other modalities.

Cross-sectional UST offers several potential clinical applications that warrant further study. This technique could enable screening of organ size or structure as an indicator of inflammation or disease [39]. For instance, liver cirrhosis may be visualized and tracked over time to monitor its progression. Aortic aneurysms may also be visualized with UST in patients who are often asymptomatic. The quantitative measurements of speed of sound and attenuation coefficient could serve as diagnostic tools, for instance, to assess changes due to non-alcoholic fatty liver disease [39]. Additionally, our speed of sound maps could enhance acoustic focusing in therapeutic applications such as shockwave lithotripsy for kidney or gallbladder stones.

Compared with other emerging techniques like low-field MRI [40], whole cross-sectional UST is faster (~10 seconds per 2D slice) with comparable or finer resolution (~1 mm), and it does not require a shielded room or magnet-compatible environment. Further, it is more portable, more open, and less noisy than MRI. Also, due to its magnet-free operation, it can be used for subjects with implants that are incompatible with MRI. When combined, these features make cross-sectional ultrasound tomography a potential practical tool for clinical needs currently unmet by other modalities. Although our system's resolution does not match clinical CT or MRI, it offers unique advantages for longitudinal monitoring by enabling frequent scans without ionizing radiation exposure, large-magnet hazards, or lengthy scanning times. The system's efficiency in both cost and time could make it practical to track changes over weeks or months, especially for conditions where broad structural changes are most relevant, such as organ size variation or adipose redistribution. We provide a comparison of cross-sectional UST with other clinical modalities in Supplementary Table 1.

Our large FOV enables adipose assessment around the entire body. Unlike calipers or probe-based ultrasound, this approach visualizes subcutaneous and preperitoneal adipose around the entire periphery without mechanical deformation. This could be an appealing tool for liposuction planning and evaluation, weight loss monitoring [41], or pharmaceutical trials for anti-obesity drugs, where MRI and CT are prohibitively costly or harmful. Improved image quality or contrast may further enable the evaluation of visceral adipose regions. Muscle regions are also observed in our images (e.g. abdominal and leg muscles), which may be useful for guiding athletic training.

UST imaging may also be appealing for monitoring liposarcoma, a malignancy in adipose tissue traditionally tracked with CT or MRI. While preliminary studies have characterized the acoustic properties of liposarcomas [42], ultrasound has not been widely adopted for treatment monitoring or surveillance. The substantial size of these tumors (typically 20 – 25 cm at diagnosis [43]) makes them potentially suitable targets for UST visualization, but we will need to validate our approach in a wider variety of patient sizes. The safety and accessibility of UST could enable more frequent monitoring (e.g., weekly assessments) compared to current imaging modalities, potentially offering new insights into treatment response and early detection of recurrence.

Cross-sectional UST shows promise in image-guided needle biopsy where CT imaging is conventionally used. Our human-width FOV enables biopsy needle localization relative to internal features without ionizing radiation. Unlike CT guidance, which requires iterative needle positioning and imaging, UST provides real-time feedback during insertion. This technique could also be used for localization with minimally invasive surgical robots [44], where the needle or tool tracked relative to internal features is particularly valuable during procedures that deform tissues.

Our current 10-second slice acquisition time enables imaging within a single breath hold. Two primary factors limit the acquisition time from reducing further: the mechanical scanning rate of the transmitter and the data acquisition transfer rate, which depends on the repetition rate and acquisition length. We could achieve faster scanning by incorporating a slip ring for electrical connection to the transmitter (like those used in CT systems) and a more powerful driving motor. Given the ~1 ms roundtrip acoustic propagation time within the immersion tank, the repetition rate could potentially increase from our current device limit of 180 Hz to ~1 kHz. While an acoustic array could also be used to transmit and receive signals, this approach may compromise sensitivity due to electrical switching and reduced chirp signal length and quality. Water immersion for acoustic coupling can be eliminated through several alternatives: inflatable water bags like those used in shockwave lithotripsy [45], gel or liquid standoff pads [46], or with skin-coupled conformal acoustic arrays [47]. However, water immersion may be practical for specific applications like assessing adipose distribution. For example, fitness or wellness centers could conveniently integrate immersion-based scanning where users already swim or bathe. These users might value periodic scanning sessions to track changes in adipose and muscle thickness over time, similar to how dual-energy X-ray absorptiometry is used to monitor body composition [48].

In the future, we also plan to enhance this system with additional photoacoustic and thermoacoustic contrast. Using the same acoustic receivers, these images would be inherently co-registered with our UST images to overlay optical and microwave absorption profiles. We also plan to improve our transmission-mode reconstruction quality using techniques such as full-wave inversion [49], enabling better localization of speed of sound and attenuation coefficient variations. Additional acoustic elements could also reduce image acquisition time and provide 3D imaging capability.

**Materials and Methods**

*System hardware*

We developed a custom 512-element, 60 cm diameter acoustic receiver array with 1 MHz center frequency. This geometry is scaled from similar systems for small animal or human breast photoacoustic imaging [50], [51]. We use lower acoustic frequencies than typical handheld probes or breast UST systems to enable whole cross-sectional imaging. For instance, typical acoustic attenuation of ~1 dB·cm$^{-1}$·MHz$^{-1}$ results in ~30 dB attenuation across a typical 30 cm diameter human cross-section at 1 MHz [21].

All 512 receiver array elements are 1 mm thick, 3 mm × 10 mm gold-coated piezoelectric polymer (PVDF-TrFE, PolyK Technologies LLC). We selected PVDF-TrFE for its broad bandwidth and ease of manufacturing. Its acoustic impedance (~4.2 MRayl) is also more closely matched to casting epoxy backings (~3.5 MRayl) and water (~1.5 MRayl) than other piezoelectric materials. Each element is capacitively coupled to copper cladded polyimide electrodes by bonding with high-strength epoxy. A continuous copper cladded polyimide electrode is used for the ground reference. The electrodes are then directly connected to parallel preamplifiers implemented on custom annular printed circuit boards. The preamplifiers provide 15 dB voltage gain with 100 kΩ input impedance. Further construction details are shown in Supplementary Fig. 1.

We machined a 60 cm diameter plastic disc and used it as a mold for the inner surface of the array. The elements and preamplifiers are housed in a stainless-steel shielded enclosure, with coaxial cables for each element connected through stainless steel holding tubes. Casting epoxy serves as the backing material for each element, and we incorporate an angled back panel to reduce acoustic reverberation (Fig. 6). All channels are low-pass filtered ($f_c = 2$ MHz) and digitized (Photosound Legion) in parallel at 5 MSPS using 20 dB additional gain. The digitizers are controlled and transfer data through USB over optical fiber to reduce interference. The preamplifiers are powered by rechargeable lithium polymer batteries with a DC voltage regulator to reduce electrical noise. To account for geometrical error during manufacturing, the technique described in [52] is used to calibrate each element's position.

The gear rotation is driven with a stepper motor. An optical homing switch is used to ensure a consistent initial rotation angle. Plastic hooks are mounted on the gear to hold the transmitter cable on the gear surface during rotation (Supplementary Video 2).

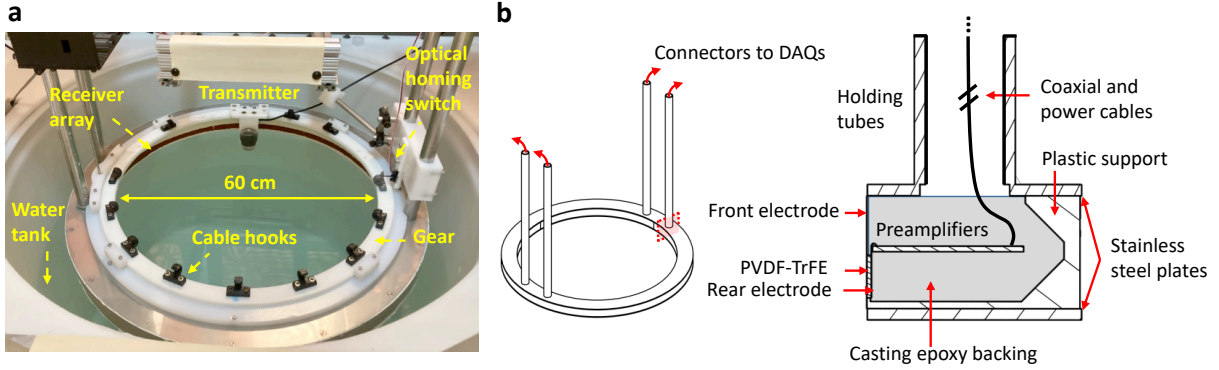

**Fig. 6. Whole cross-sectional UST hardware. a** System photograph. **b** Acoustic receiver array design, showing a cross-section of the array. DAQs: data acquisition modules.

*Acquisition parameters*

To enhance the signal-to-noise ratio (SNR) while limited by the mechanical index, a linear chirp signal versus time ($t$) is used with a time varying frequency $f(t) = f_r t + f_0$, where $f_r = (f_1 - f_0)/T$ is the linear chirp rate, $f_0 = 0.3$ MHz is the lower frequency, $f_1 = 2.0$ MHz is the upper frequency, and $T = 400$ µs is the chirp duration. The transmitted frequencies are limited by the bandwidths of the transmitter and receivers. We used a maximal pulse duration given our maximal acquisition time of 800 µs, allowing for recovery of the roundtrip reflected signals over the entire FOV. The resulting transmitted chirp signal is

$$x(t) = \sin\left[2\pi\left(\frac{f_r}{2}t^2 + f_0 t\right)\right]. \tag{1}$$

Compared to a pulse with similar peak pressure, this results in an expected SNR gain of $\sim \sqrt{T \cdot B}$, where $B = f_1 - f_0$ is the acoustic bandwidth [53]. In addition to the target, we also perform a scan with only water in the imaging domain, resulting in recorded signals $x_{w,i}(t)$ for each receiver element $i$. This provides the response of each transducer to the chirp which is then cross-correlated with the target's chirp response $x_{c,i}(t)$. The pulse response for the target signals $\chi_{s,i}(t)$ is then recovered for each element $i$ as:

$$\chi_{s,i}(t) = \frac{x_{w,i}(t) \star x_{c,i}(t)}{\max[x_{w,i}(t) \star x_{w,i}(t)]}, \tag{2}$$

where $\star$ denotes cross-correlation. We normalize by the maximum of the autocorrelation of $x_{w,i}(t)$ to account for sensitivity variation in the receiver elements. The transmitter operates with a pulse repetition rate of 180 Hz. With the gear rotation time of 10 seconds, this results in 1800 transmitted pulses over a full circular scan around the target.

*Image quality*

We assessed our in-plane resolution using a thin (< 0.1 mm) metallic wire. The reconstructed image and axis profiles are given in Fig. **7**a and b. We find an in-plane FWHM of approximately 0.9 mm. To determine our elevational resolution, we imaged a thin brass disc positioned such that its edges were at the center and outer boundary of our typical FOV (Fig. **7**c). The height of the disc was scanned, and a UST image was obtained at 2.5 mm increments. An elevational FWHM of 15 mm and 25 mm was found for the center and edge of the FOV, respectively. Neither the transmitter nor receivers are focused in the elevational direction, but their larger dimensions in this direction reduces their acceptance angle.

*Human imaging protocol*

Four (three female, one male) volunteers consented to be imaged in this system. This imaging procedure was approved by the Caltech Institutional Review Board (protocol IR21-1099). Prior to human imaging, we used a calibrated hydrophone (Onda HGL-0085) positioned immediately in front of the transmitter to evaluate the safety of our system. We calculated a mechanical index as less than 0.2 over the entire chirp bandwidth, whereas the limit from the U.S. Food and Drug Administration (FDA) is 1.9 [54]. We evaluated the spatial peak temporal average intensity ($I_{SPTA}$) as 24 mW/cm$^2$, whereas the FDA limit is 720 mW/cm$^2$. We calculated the thermal index from the empirical formula as listed in [55] to be 0.65, which is also much lower than the FDA limit of 6.0 for diagnostic ultrasound devices.

Prior to imaging, we filled the water immersion tank with warm (30 °C) tap water using rubber tubing. We provided a private room for volunteers to change into a swimsuit and to dry using a towel and space heater after the imaging session. Volunteers entered the tank using an external ladder with handrails and anti-slip coatings (Supplementary Fig. 11). After volunteers entered the water tank, we asked them to use their hand to wipe away air bubbles that may have accumulated on their abdomens. An emergency off button was accessible to the volunteers during scanning, shutting electrical power to all system devices. All electrical devices were powered through a ground-fault circuit interrupter in case of water exposure. The operators ensured volunteer comfort verbally and visually. The volunteers remained seated on an immersed stool during image acquisition with their shoulders resting on a backrest to reduce motion. We performed these human experiments in a dedicated imaging room. We obtained written informed consent from participants according to our IRB protocol.

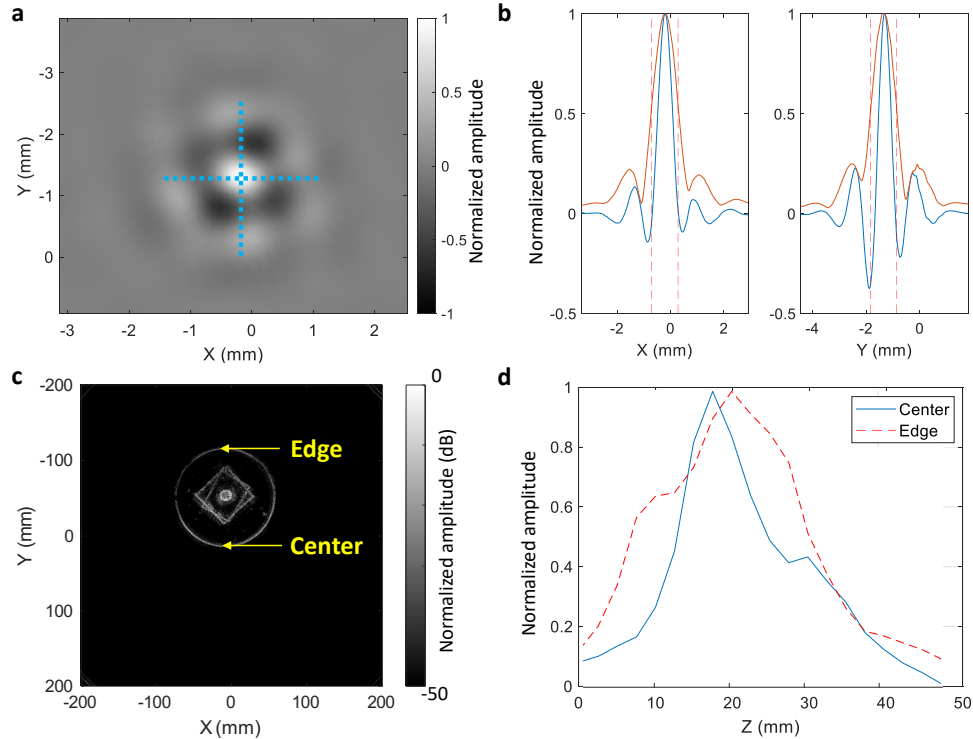

**Fig. 7. In-plane and elevational resolution assessment. a** Reconstructed reflectivity image of a thin wire. **b** Profiles along the x and y axes: pixel amplitude (blue), the magnitude of its Hilbert transform (orange), and the full width at half maximum (FWHM, dashed vertical lines). **c** Example reflectivity image of a brass disc used for elevational resolution assessment. **d** Profiles of the center and edge responses at different z positions.

## Acknowledgements

We are grateful to Dr. William Tseng for his valuable feedback and insightful suggestions regarding future research directions. This work was supported in part by National Institutes of Health grants R35 CA220436 (Outstanding Investigator Award) and by grant number 2024-337784 from the Chan Zuckerberg Initiative DAF, an advised fund of the Silicon Valley Community Foundation. L.W. has a financial interest in Microphotoacoustics, Inc., CalPACT, LLC, and Union Photoacoustic Technologies, Ltd., which, however, did not support this work.

## Data and code availability

The data that support the findings of this study are provided within the paper and its Supplementary materials. The reconstruction algorithm and data processing methods can be found in the paper. The reconstruction code is not publicly available because it is proprietary and may be used in licensed technologies.

**Competing interests:**

**Supplementary Information**

# Whole Cross-Sectional Human Ultrasound Tomography

David C. Garrett[†], Jinhua Xu[†], Yousuf Aborahama, Geng Ku,
Konstantin Maslov, and Lihong V. Wang*

*Caltech Optical Imaging Laboratory, Andrew and Peggy Cherng Department of Medical Engineering, Department of Electrical Engineering, California Institute of Technology, Pasadena, CA 91125 USA*

[†] These authors contributed equally.
* Corresponding author: lvw@caltech.edu

*Comparison to other modalities*

We compare cross-sectional UST to other diagnostic modalities in Supplementary Table 1. While the commercial cost of our approach is not yet known, we expect it to be low given the relatively small (< $100k) material and equipment cost we used to develop our system.

Supplementary Table 1. Comparison of abdominal imaging modalities.

|  | *X-ray CT* | *MRI* | *Probe-based ultrasound* | *Cross-sectional UST* |
|---|---|---|---|---|
| *Whole cross-section penetration (~35 cm)* | Yes | Yes | No | Yes |
| *Whole abdominal imaging time* | Moderate (5 – 10 min.) | Very long (30 – 60 min.) | Moderate overall time despite very fast individual images (10 – 20 min.) | Moderate (5 – 10 min.) |
| *Safety* | Ionizing radiation (~5 – 15 mSv) | Strong magnetic fields, radiofrequency exposure | Safe | Safe |
| *In-plane resolution* | Very good (~0.5 – 1 mm) | Good (~1 – 2 mm) | Very good, scalable (~0.3 – 2 mm) | Good (~1 mm) |
| *System cost* | Moderate | Very high | Low | Low |
| *Operational cost* | Moderate (Maintenance, X-ray tubes) | Very high (Technicians, liquid helium) | Moderate (Trained operation) | Low (Water refilling, minimal operator training) |

We compare cross-sectional UST to conventional methods of adipose assessment in Supplementary Table 2. Note that calipers, probe-based ultrasound, and our approach estimate subcutaneous adipose (SA) thickness, whereas DXA estimates body composition in 2D projections.

Supplementary Table 2. Comparison of adipose assessment modalities. DXA: dual-energy X-ray absorptiometry. SA: subcutaneous adipose.

|  | *Calipers* | *DXA* | *Probe-based ultrasound* | *Cross-sectional UST* |
|---|---|---|---|---|
| *Measured property* | Thickness of a pinched fold of skin and SA | Dual energy X-ray absorption | Acoustic backscatter | Acoustic backscatter |
| *Estimated value* | SA thickness | Fat content instead of thickness | SA thickness | SA thickness |
| *Safety and comfort* | Slight pinch on skin, procedural intimacy | Low ionizing radiation exposure (~1 – 10 μSv)[*] | Procedural intimacy | Water immersion |
| *Assessed region* | Individual superficial sites | Whole body in 2D projections | Individual sites | Whole abdomen and lower body |
| *Operator dependence* | High (Location selection, mechanical deformation) | Very low, but strong dependence on calibration | High (Location selection, mechanical deformation) | Low |
| *System cost* | Very low | Low | Low | Low |
| *Operational cost* | Moderate (Trained operation) | Low (Minimal operator training) | Moderate (Trained operation) | Low (Water refilling, minimal operator training) |

[*]Despite the low ionizing radiation exposure, some patients may prefer to avoid any additional exposure. The recommended annual radiological exposure limit is 1 mSv [1].

We compare approaches for biopsy needle localization in Supplementary Table 3.

Supplementary Table 3. Comparison of modalities for biopsy needle localization.

|  | *X-ray CT* | *Probe-based ultrasound* | *Cross-sectional UST* |
|---|---|---|---|
| *Feedback time* | Iterative, long (~Minutes) | Real-time | Real-time |
| *Operational complexity* | Minimal | High (Manipulating the probe and needle simultaneously) | Minimal |
| *Safety* | Ionizing radiation (~10 mSv) | Minimal | Minimal |
| *Field of view* | Whole cross-section | Local to the probe | Whole cross-section |
| *Anatomical sites* | Any organ | With ultrasound window | With ultrasound window |

*Array construction*

We show the steps of the array construction in Supplementary Fig. 1. The 1 mm thick, gold-coated PVDF-TrFE piezoelectric was diced into elements 3 mm in width and 10 mm in height using an excimer UV laser. The elements were then aligned with copper-clad polyimide electrodes and bonded using high-strength epoxy, leading to capacitive coupling. We extended a short electrical trace from each electrode for connection to the preamplifier circuits. We then bonded a continuous piece of copper-clad polyimide on the opposite surface of the piezoelectric material.

Strips of 128 channels were then mounted into the plastic supports and positioned around a 60 cm diameter lathed disc of high-density polyethylene to provide a precise template surface geometry. The backing was then filled with casting epoxy. Next, the preamplifier printed circuit boards (PCBs) were mounted, and each electrode trace was soldered to its amplifier channel. Every ~10 channels, a short wire was also soldered to the reference (front) electrode. Signal coaxial cables for each channel and DC power cables extend down the supporting tubes and connect to the PCBs. The other sides of the cables connect to the DAQs using Acuson-style connectors mounted to custom low-pass filter PCBs. We then lowered the top stainless-steel plate and trimmed the front electrode to the plate height. Finally, we glued a stainless-steel shim with silver epoxy to the exterior surface of the array to provide electrical shielding.

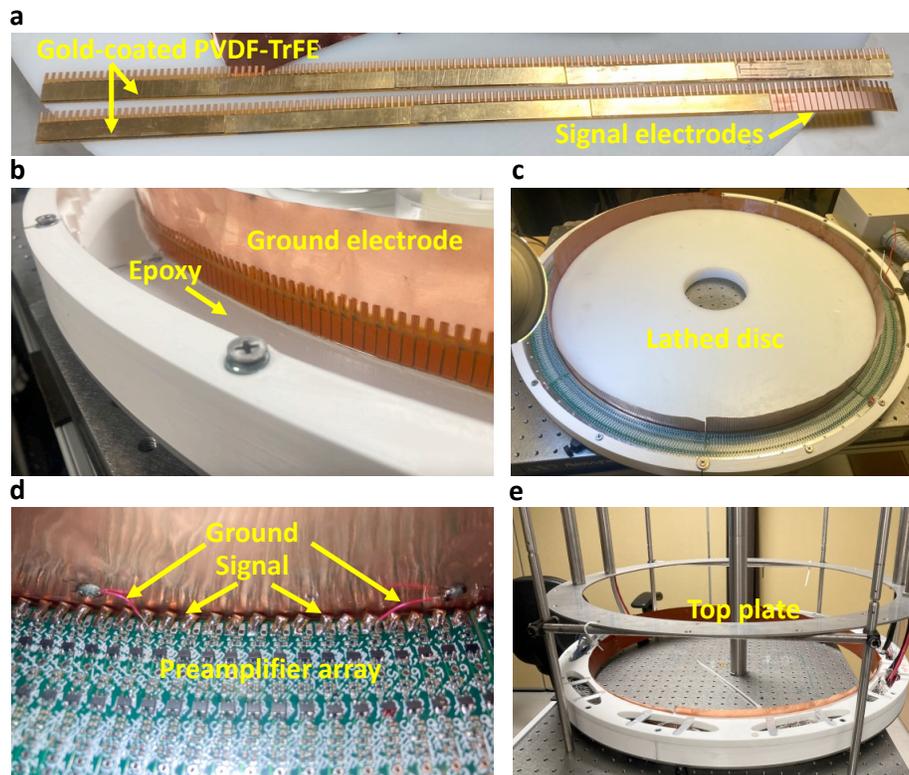

**Supplementary Fig. 1. Steps of the acoustic receiver array construction. a** Diced PVDF-TrFE is glued to polyimide electrodes and a continuous common electrode. **b** Sections of channels are mounted to a circular template. A casting epoxy backing is poured. **c** Annular preamplifier PCBs are mounted on top of each section, and the electrodes are soldered to them. **d** A grounding wire is soldered every ~10 channels. **e** The outer electrode is trimmed, the array is filled with casting epoxy, and the top stainless steel plate is screwed in.

*Receiver characterization*

The bandwidth of the array elements was evaluated using a photoacoustic point source made from carbon powder-epoxy mixture on the tip of an optical fiber (Thorlabs FT600EMT) of 600 μm core size. We used a sampling rate of 20 Msamples per second to record the bandwidth up to 10 MHz. We show the frequency response of the array elements in Supplementary Fig. 2. The average –3 dB bandwidth across the array is approximately 0.81 MHz (corresponding to 81% fractional bandwidth).

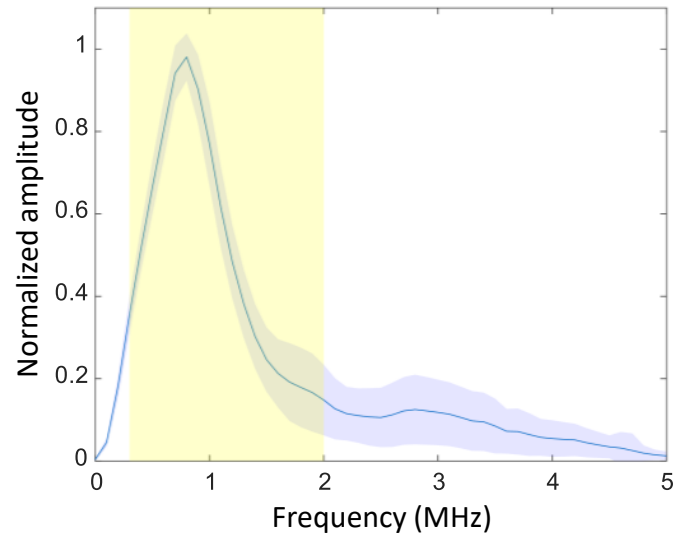

**Supplementary Fig. 2. Normalized frequency response the receiver array**. The solid curve indicates the mean response across 512 elements. The blue shaded area indicates one standard deviation from the mean response. The yellow shaded area indicates the frequency range of 0.3 – 2 MHz of the imaging chirp signal.

*Additional human images*

We also imaged the female subject from Fig. 2 at various heights from the ribcage toward the pelvis, shown in Supplementary Fig. 3. This subject had her left kidney removed during childhood but is now healthy.

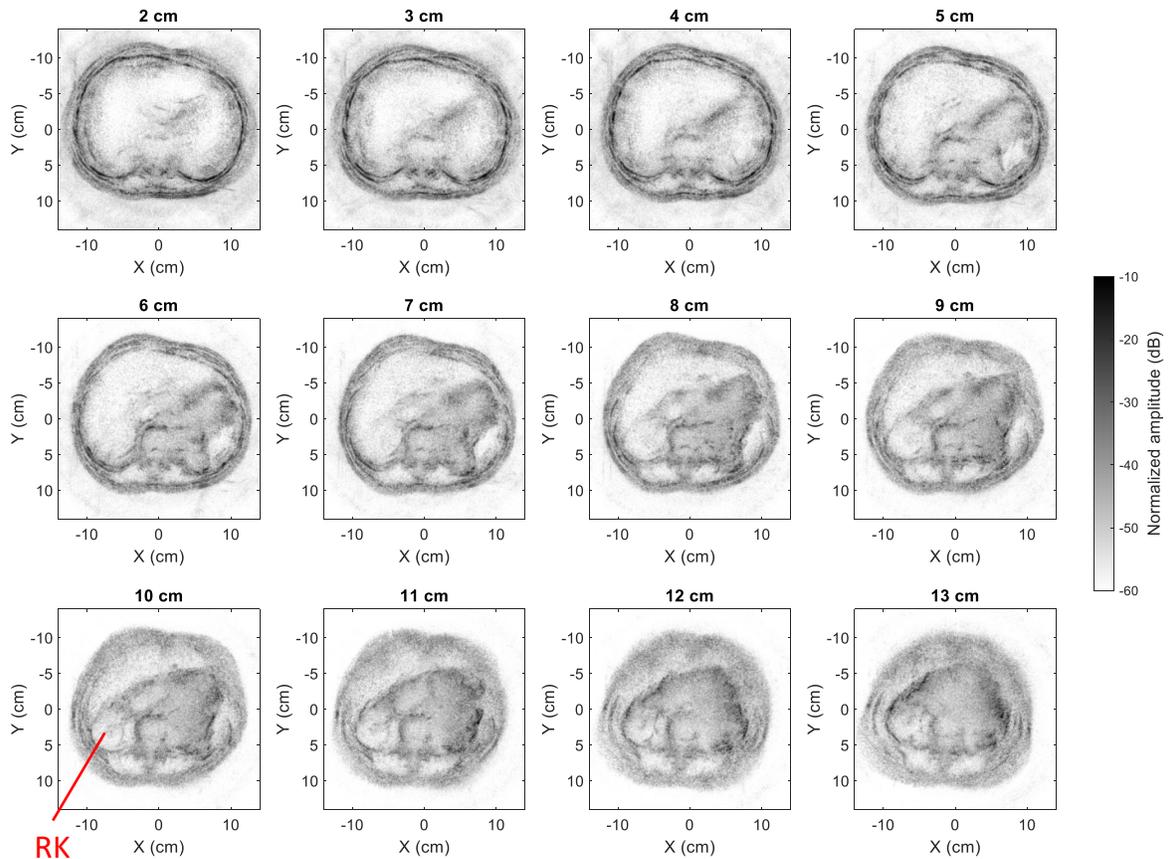

**Supplementary Fig. 3. Example of elevational scans of a female subject from approximately the ribcage to the pelvis.** RK: right kidney (left kidney was removed).

*Reflection-mode reconstruction algorithm*

We employ a 2D delay-and-sum algorithm for reflection-mode ultrasound reconstruction. We first cross-correlate each channel's recorded signal with its direct transmission water-only response to obtain a pulse-like representation. We then band-pass filter the cross-correlated signals within 0.3 – 2.0 MHz and normalize them based on the water-only autocorrelation. Then, for each transmitter position, we determine the time delay between the transmitter focal point and each pixel for a given background speed of sound. We mathematically correct for the angle-dependent focal point with the acoustic lens, as shown in Supplementary Fig. 8. Next, we calculate the delay time from each pixel to each receiver position. Each final pixel value is found from the sum of the cross-correlated received signals at the total round-trip delay time. For each transmitter position, we use the 256 receiver channels on the side of the transmitter to limit reconstruction to back-scattered signals.

We implemented this algorithm using CUDA kernel called from MATLAB, computed using a NVIDIA GeForce RTX 3070 graphics processing unit (GPU). Reconstruction was performed over a 512 mm x 512 mm grid with 0.125 mm x 0.125 mm pixel size, consisting of 4096 x 4096 pixels. For each slice, the required computational time was approximately 12 minutes. We expect this could be greatly shortened with more capable GPUs. To visualize the images, we first used a 2D Wiener filter with a 4-pixel neighborhood size. We then normalized the images to their maximum amplitude and displayed them in a logarithmic scale. Background segmentation was done using an automated segmentation tool [2].

*Straight-ray transmission-mode reconstruction algorithms*

Our data processing steps for transmission-mode imaging are shown in Supplementary Fig. 4. Starting from the recorded water-only and target signals, we first perform cross-correlation to obtain pulse-like representations of the transmitted signals. We then determine the arrival time of the water and target signals, and we determine the arrival time difference for each receiver and transmitter position pair. This is used for reconstruction of the speed of sound. Similarly, we also determine the amplitude ratio between the water and target signals, which is used for attenuation coefficient reconstruction.

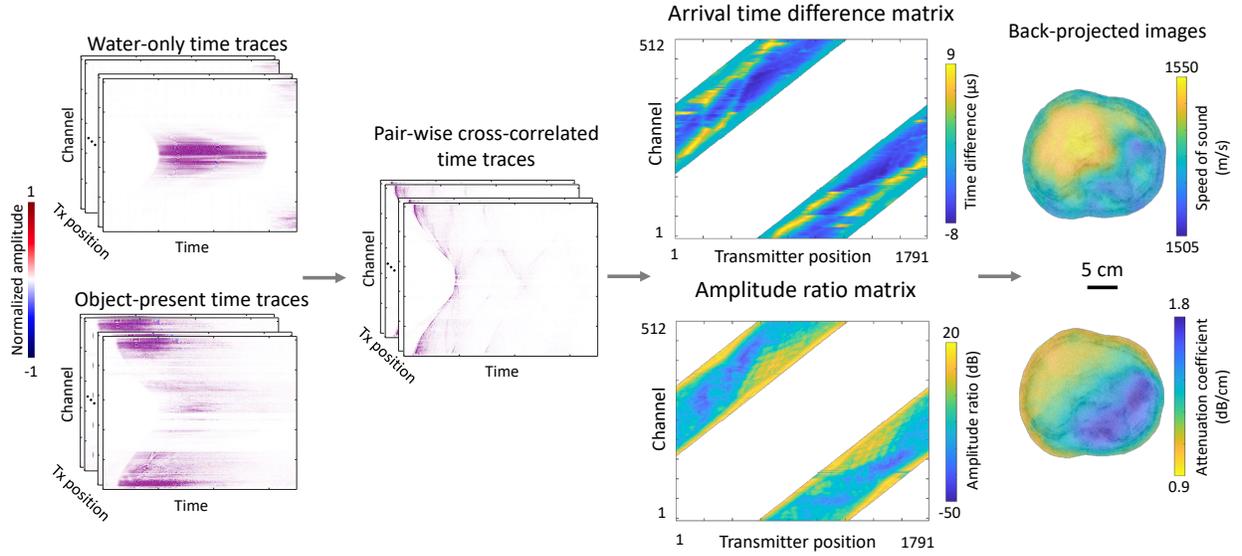

**Supplementary Fig. 4. Signal processing flow for transmission-mode analysis.**

We use the first-break method [3] to determine signal arrival time. For a given cross-correlated target signal $\chi_s(i)$ at discrete time samples $i$, an energy ratio er is defined for a window size $N_e$ as:

$$\text{er}(i) = \frac{\sum_{j=i}^{i+N_e} \chi_s^{\,2}(j)}{\sum_{j=i-N_e}^{i} \chi_s^{\,2}(j)}. \tag{3}$$

We use a window $N_e$ of 4 μs as a compromise between robustness and precision. The modified energy ratio (mer) is then calculated as:

$$\text{mer}(i) = (|\chi_s(i)| \times \text{er})^3. \tag{4}$$

The signal arrival time is determined from the maximum value of mer. As an example, the calculated transmission-mode arrival times through water and a human abdomen are shown in Supplementary Fig. 5.

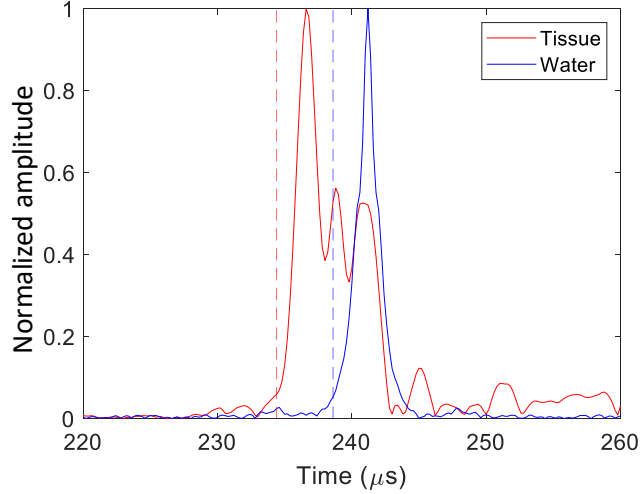

**Supplementary Fig. 5. Example cross-correlated transmission signals for water-only and the human abdomen.** The vertical dashed lines show the estimated arrival times using the first-break method.

We then perform reconstruction as follows. Along the straight ray $L_i$ between the $i^{\text{th}}$ transmitter-receiver pair, the total delay $\tau_i$ can be modeled as $\int_{L_i} \frac{dl}{c(\vec{r})} = \tau_i$, where $c(\vec{r})$ represents the location-dependent speed of sound. This equation can be discretized as $\sum_{j \in \{L_i\}} \Delta l_{ij} \cdot x_j = \tau_i$, where $x_j = \frac{1}{c_j}$ is the inverse of the speed of sound at the $j^{\text{th}}$ pixel along ray $L_i$ and $\Delta l_{ij}$ represents the length of ray $L_i$ crossing the $j^{\text{th}}$ pixel. The linear equation can be arranged into the matrix form $AX = T$, where $X$ and $T$ are the vectorized $x_j$ and $\tau_i$.

One technical issue with solving the linear equation is with the memory required for matrix $A$ which is of the size $(N_t \cdot N_r) \times (N_x \cdot N_y)$ (>$10^{11}$ in our implementation). Rather than saving and inverting the matrix directly, we take advantage of the fact that $A' = A^T A$ is close to diagonal (as neighboring rays go through neighboring pixels), and therefore $X' \approx A^T A X$ can be estimated as a vector with elements $x'_j \approx A'_{jj} x_j = \left( \sum_i \Delta l_{ij}^2 \right) \cdot x_j$. The right-hand side of the equation is then converted to $B = A^T T$, with elements $b_j = \sum_i \tau_i \Delta l_{ij}$. Thereby, simple element-wise division of $B$ by $\sum_i \Delta l_{ij}^2$ solves for $X$.

The attenuation coefficient map can be rearranged as $\int_{L_i} \mu_a(\vec{r}) dl = a_i \Rightarrow \sum_{j \in \{L_i\}} \Delta l_{ij} \cdot \mu_{a\,j} = a_i$ and solved concurrently.

We also tested methods such as global path tracing [4], an adapted algorithm of dynamic time warping used in seismology field, to enforce the smoothness prior between neighboring frames, and matched filtering which requires accurate knowledge of the transmitted waveform. In practice, we found these methods to perform similarly given the SNR of the time traces collected.

*Transmission-mode reconstruction accuracy*

To assess the accuracy of our speed of sound profiles, we imaged ethanol-water mixtures with speed of sound ranging from approximately 1510 to 1610 m/s. Note that the speed of sound increases with ethanol content until ~50%, but decreases at higher content [5]. The reference speed of sound was determined using a single-element transducer immersed in the mixture and by varying the distance to an acoustic reflector to record the time-of-flight using an ultrasonic pulser (Olympus 5072PR). Measurements were recorded over 21 distances at 1 mm intervals. The speed of sound was then calculated as $v = \Delta d / 2\Delta t$, where $\Delta d$ is the difference in position, and $\Delta t$ is the difference in arrival time. The arrival time is determined using the first-break approach. Supplementary Table 4 shows the mean and standard error of the speed of sound estimations for each mixture.

We then held the five mixtures in thin membranes with ~3 cm diameter and imaged together in the same session. For transmission-mode reconstruction, we used both the unmasked and masked approaches, where the masked method uses the reflection-mode image to segment each mixture during reconstruction. The resulting speed of sound estimations are given in Supplementary Table 4. We find typical errors of less than 1%, but a maximum error of 1.52% and 2.84% for the unmasked and masked approaches, respectively.

Supplementary Table 4. Estimated speed of sound in m/s for ethanol-water mixtures. Reference values were determined using a single-element transducer at varying distances in the mixtures. Percent error from the reconstructed values is shown in parentheses.

| Method | Ethanol content | | | | |
| --- | --- | --- | --- | --- | --- |
| | **5%** | **10%** | **20%** | **50%** | **60%** |
| Reference | 1508.6 | 1532.4 | 1578.9 | 1609.5 | 1531.4 |
| (Standard error) | (0.18) | (0.24) | (0.26) | (0.50) | (0.18) |
| Unmasked | 1511.0 | 1524.5 | 1580.2 | 1604.2 | 1508.1 |
| (% error) | (0.15%) | (-0.52%) | (0.08%) | (-0.33%) | (-1.52%) |
| Masked | 1509.39 | 1542.04 | 1578.83 | 1563.77 | 1520.03 |
| (% error) | (0.04%) | (0.63%) | (0.00%) | (-2.84%) | (-0.74%) |

From our human images, we then compared our estimates of several tissues' speed of sound (from the subject in Fig. 2) with literature values [6], as shown in Supplementary Fig. 6. Our estimates for the liver and stomach speed of sound match closely with literature values, but there is less agreement for other organs like the spleen. This is likely because these regions are smaller and therefore undergo less temporal shift from their speed of sound variations.

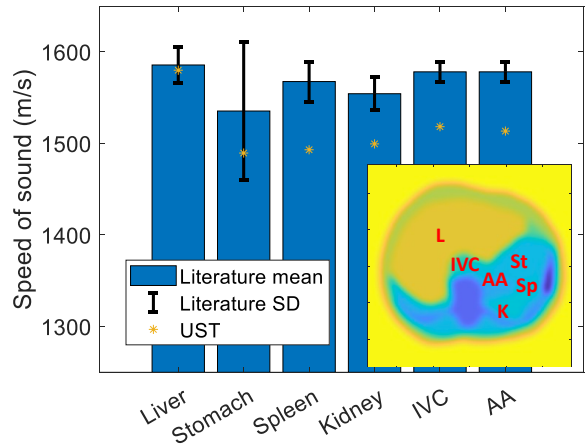

**Supplementary Fig. 6. Comparison of literature [6] and UST estimations of tissue speed of sound.** SD: standard deviation. L: liver. St: stomach. Sp: spleen. K: kidney. IVC: inferior vena cava. AA: abdominal aorta. The inset shows the segmented tissue regions used for speed of sound reconstruction.

Since liver health assessment is a potential application of cross-sectional UST, we compared our estimated properties with reported values in various diseased states. Supplementary Fig. 7 shows the reported speed of sound [7] and attenuation coefficient [8] of the healthy and diseased liver. These values are shown alongside our estimations (from the subject in Fig. 2) across six image slices spaced by 1 cm vertically. As expected, our estimated values fall within the expected range for healthy subjects.

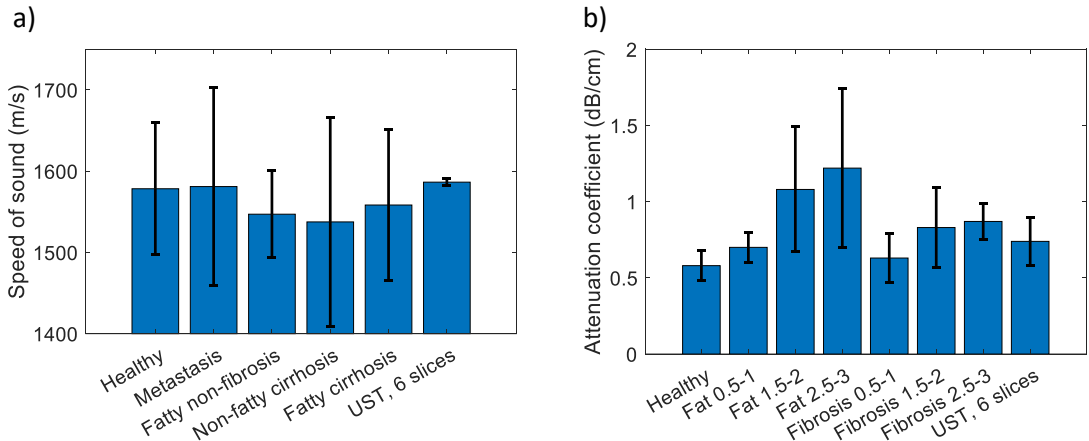

**Supplementary Fig. 7. Comparison of UST transmission-mode estimations for a healthy subject's liver acoustic properties.**
**a** Literature speed of sound values for the liver under various diseased conditions [7]. **b** Literature attenuation coefficient values for the liver at varying pathological stages of disease [8]. Error bars show the standard deviation of literature values and UST estimations across six images.

*Transmitter lens correction*

For manufacturing simplicity, we used a cylindrical TPX lens to diverge the transmitter beam, shown in Supplementary Fig. 8a. If uncorrected, this geometry causes cylindrical aberration where different emission angles correspond to varying effective focal lengths. We correct for this aberration during reconstruction. For each transmitter position, we determine the angle to each pixel and assign an appropriate focal length. We consider a plane wave emitted from the flat transducer surface as shown in Supplementary Fig. 8b. Wavefronts emerging at position $x$ along the transducer surface arrive at the lens interface at depth $y = \sqrt{r_{lens}^2 - x^2}$. The angle normal to the lens surface at this position is:

$$\theta_{norm} = \tan^{-1}(x/y). \qquad (5)$$

The wavefront refracts at an angle relative to the normal angle as:

$$\theta_{ref} = \sin^{-1}\left[\sin(\theta_{norm})\frac{c_{water}}{c_{lens}}\right], \qquad (6)$$

where $c_{water} \sim 1500$ m/s and $c_{lens} \sim 2090$ m/s are the water and TPX lens speed of sound, respectively. The angle into the imaging plane is therefore $\theta_{pix} = \theta_{norm} - \theta_{ref}$. This line intersects with the centerline of the transducer at distance $f$ behind the emitting surface:

$$f = \frac{x}{\tan \theta_{pix}} - y. \qquad (7)$$

During reconstruction, from a given $\theta_{pix}$, we can therefore calculate the appropriate focal length $f$. Example calculated values are shown in Supplementary Fig. 8c. This is incorporated in our GPU-based reconstruction.

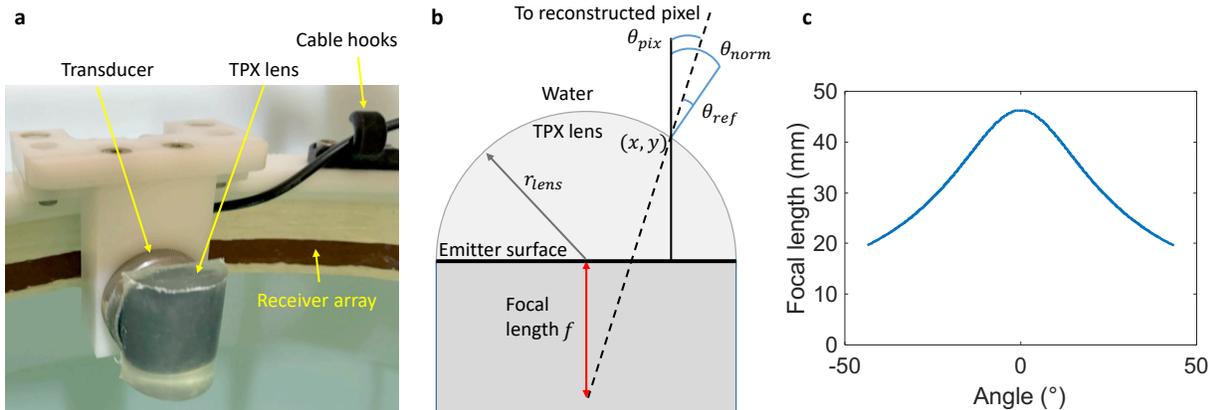

**Supplementary Fig. 8. Determining the effective focal length for various pixel angles using a cylindrical TPX lens on the acoustic transmitter. a** Photograph of TPX lens mounted on the transmitting transducer. **b** Diagram of TPX lens and transmitter surface. **c** Calculated focal length for varying angles to reconstructed pixels ($\theta_{pix}$).

*Biopsy needle acoustic modes*

The biopsy needle acts as an acoustic waveguide where different modes are supported depending on the frequency and form of excitation. These modes can be longitudinal, torsional, or flexural. We expect the dominant mode to be the longitudinal L(0,1) mode which, fortunately, has approximately uniform phase velocity and low attenuation over our frequency range for this needle diameter (~1.5 mm), as shown in Supplementary Fig. 9a and b [9]. The phase velocity is close to the bulk longitudinal velocity of stainless steel (~5790 m/s). During reconstruction, we therefore consider a constant propagation delay down the needle length of ~50 μs.

An example received waveform spectrogram is shown in Supplementary Fig. 9c, showing that the linear chirp is maintained after propagation along the needle and emission at the tip. There is some reverberation after the dominant chirp signal due to other acoustic modes. A weak second-order chirp (at twice the dominant frequencies) is also visible due to nonlinearities in the power amplifier.

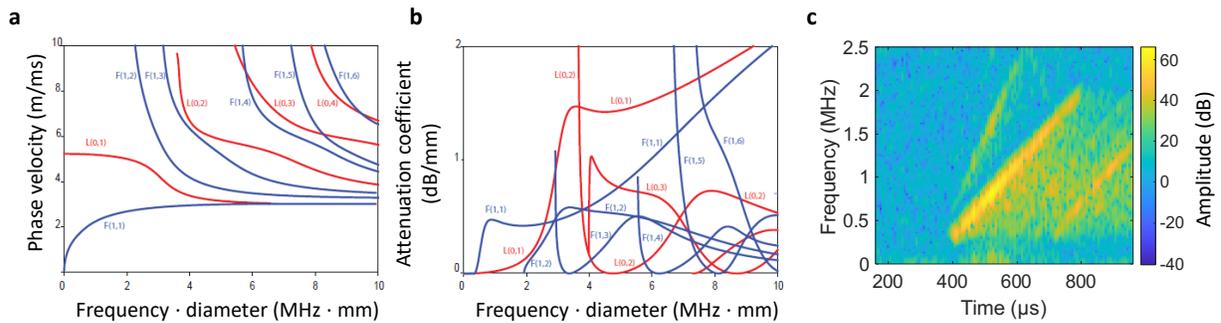

**Supplementary Fig. 9. Expected acoustic modes in an unmodified biopsy needle. a** and **b**: Phase velocity and attenuation coefficient, respectively, for acoustic modes supported by a stainless steel rod immersed in water, adapted from [9]. **c** Example waveform received from the scattered emission from the biopsy needle.

*Adipose thickness estimation*

To validate our adipose thickness estimation method, we constructed a phantom consisting of a lard layer over a 4% agar core (Supplementary Fig. 10a and b). A plastic ribbon of 0.25 mm thickness defines the material boundaries. Ridges in plastic plates on the top and bottom of the phantom align the ribbon in precise shapes. Heated liquid agar was first poured in the center mold and allowed to cure at room temperature for two hours. We then melted pork lard and poured it into the outer mold. The phantom then cooled at room temperature for one hour. The lard thickness on two sides of the phantom is 1.0 and 2.0 cm, and the phantom is 7.5 cm tall.

We then imaged the phantom in the UST system (Supplementary Fig. 10d). To estimate the lard layer thickness, we first extract line profiles from the reflection-mode images and plot the image amplitude $A$ versus position $s$ along the line. Supplementary Fig. 10e and f show the normalized image amplitude along lines on two sides of the phantom. We estimate the layer thickness by determining the two dominant amplitude peaks (at positions $s_{0,1}$ and $s_{0,2}$) and calculating the distance between them.

We estimate the adipose layer thickness uncertainty $\sigma_t$ as

$$\sigma_t = \sqrt{\frac{\sigma_{w1}^2}{\text{CNR}_1} + \frac{\sigma_{w2}^2}{\text{CNR}_2}}, \tag{8}$$

where $\sigma_{w1}$ and $\sigma_{w2}$ are the root mean square widths of the amplitude peaks, and $\text{CNR}_1$ and $\text{CNR}_2$ are their amplitude-based contrast-to-noise ratio (CNR). Since the peaks do not necessarily follow Gaussian shapes, we calculate the widths for each peak $i$ as:

$$\sigma_{w,i} = \sqrt{\int_{s_{0,i}-s_{win}}^{s_{0,i}+s_{win}} (s - s_{0,i})^2 \cdot A(s)ds \bigg/ \int_{s_{0,i}-s_{win}}^{s_{0,i}+s_{win}} A(s)ds}. \tag{9}$$

We consider a window $s_{win}$ around each side of the peaks (chosen as 5 mm) to capture the dominant interface response without extending to neighboring features. We calculate the CNR in a linear scale as

$$\text{CNR}_i = \frac{A(s_{0,i}) - \overline{A_{out}}}{\text{std}(A_{out})}. \tag{10}$$

Here, $A_{out}$ is a 4 cm portion of the line profile outside the target used to characterize the background.

The estimated thicknesses of $1.05 \pm 0.02$ cm and $2.02 \pm 0.05$ cm agree closely with the true values of 1.00 cm and 2.00 cm, respectively. A ~3% overestimation in thickness is expected due to the slower speed of sound in lard (~1440 m/s) than room temperature water (~1483 m/s), which we do not correct for here. We used the same methodology to estimate the adipose thickness in our human images in Fig. 4.

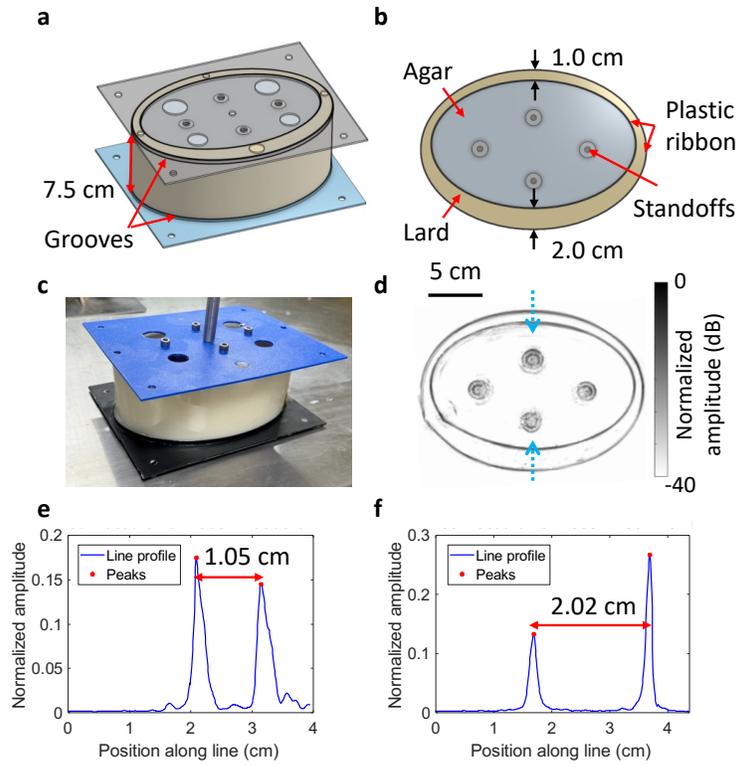

**Supplementary Fig. 10. Estimating adipose thickness in a phantom. a** Phantom design. A plastic ribbon holds a layer of pork lard around an agar core. The lard thickness on two sides is to be 1.0 cm and 2.0 cm. **b** Reflection-mode UST image of the adipose phantom. Dashed lines show the lines used for extracting the lard layer thickness, with the arrow showing the direction of the line profile. Panels **c** and **d** show the normalized amplitude (in linear scale) extracted from the drawn lines in the UST image on the 1.0 cm and 2.0 cm lard thickness sides of the phantom, respectively.

*System construction*

We constructed the primary structure of our system using 1.5-inch aluminum T-slots. A high-density polyethylene plate was water jet cut with openings for the water tank and linear stages. An aluminum plate was water jet cut and mounted on the entry side of the tank, and an anti-slip polyurethane mat was adhered on its surface. A dock ladder is mounted to the aluminum plate with handrails and anti-slip steps for safe entry into the immersion tank (Supplementary Fig. 11).

The larger gear was water jet cut from acetal with 23.5-inch outer diameter and 564 gear teeth. This gear was positioned in an ultra-high molecular weight polyethylene mount with a lathed rotational surface. The mount was screwed to the top plate of the acoustic receiver array. We used an 18-tooth brass pinion gear to drive the larger gear's rotation, coupled using a rotary shaft to a stepper motor. The resulting gear ratio is ~31.3. The gear is held into the mount using plastic pieces screwed into the gear mount. The transmitter is held to the gear using a magnetic mount for free release in case of obstruction by the volunteer.

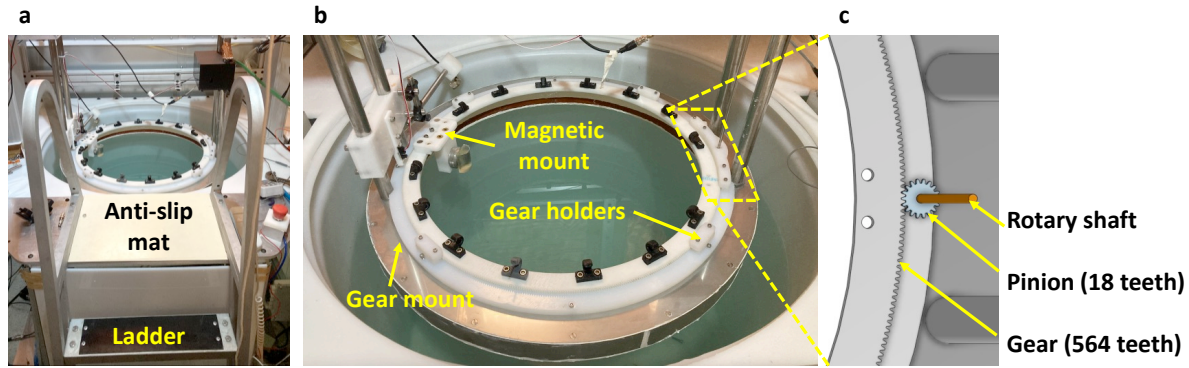

**Supplementary Fig. 11. Additional details of the UST system construction. a** Immersion tank entry using a ladder with handrails and an anti-slip mat. **b** Top side of the receiver array, showing the gear and its mount and holding pieces. **c** Model of the gear driver, with the rotary shaft connected to a stepper motor.

To avoid water immersion in the future, we plan to adapt our system using filled water membranes that couple acoustic signals between the array elements and the skin. We show an example configuration in Supplementary Fig. 12. Elastic membranes could accommodate a variety of body sizes using different amounts of filled water. Ultrasound gel on the skin surface may further improve acoustic coupling.

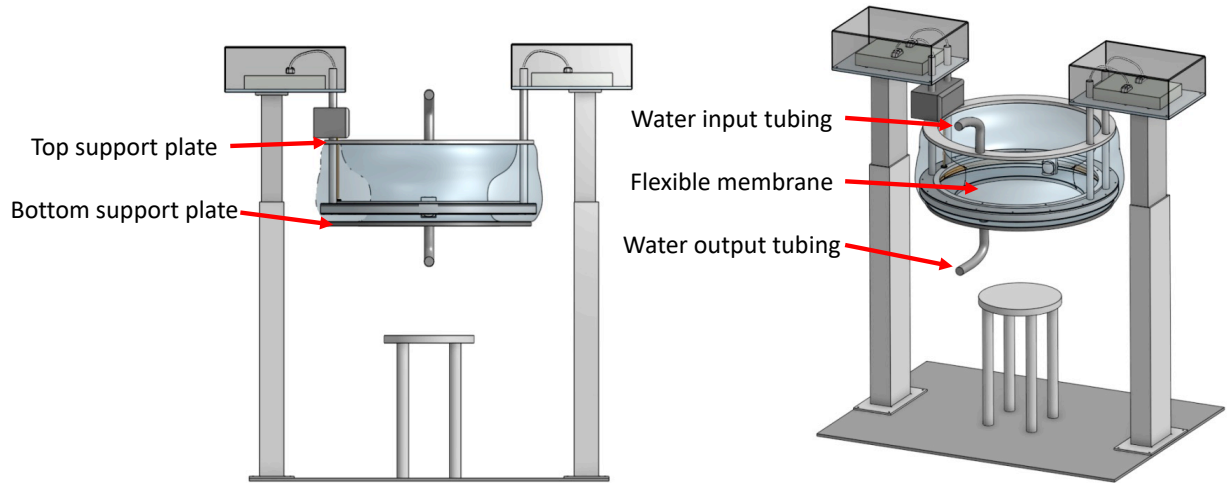

**Supplementary Fig. 12. Potential adaptation of our system to enable immersion-free abdominal imaging.**